\documentclass[12pt]{article}
\usepackage{amsmath}
\usepackage{graphicx}
\usepackage{enumerate}
\usepackage{natbib}
\usepackage{url} 

\newcommand{\blind}{1}

\usepackage{caption,subcaption,graphicx,psfrag,epsf,cases,empheq,color,amssymb,amsmath,graphicx,float,booktabs,amsthm,bm,bbm,multirow,threeparttable,textcomp,adjustbox,setspace,natbib,amsthm,soul,cancel,mathrsfs,euscript,amsfonts,xcolor,tikz,bm,blkarray,stackrel,ulem}
\usepackage[hidelinks]{hyperref}
\theoremstyle{plain}

\newtheorem{theorem}{Theorem}

\newtheorem{remark}{Remark}
\newtheorem{example}{Example}
\newtheorem{assumption}{Assumption}

\newtheorem*{assumption*}{\assumptionnumber}
\providecommand{\assumptionnumber}{}


\providecommand{\assumptionnumber}{}


\providecommand{\assumptionnumber}{}
\makeatletter
\newenvironment{confoundingbridge}[1]
{
	\renewcommand{\assumptionnumber}{Assumption #1$'${\normalfont~(Existence of confounding bridge)}}
	\begin{assumption*}
		\protected@edef\@currentlabel{#1$'$}
	}
	{
	\end{assumption*}
}
\makeatother

\newcommand{\ken}[1]{\textcolor{black}{#1}}

\providecommand{\assumptionnumber}{}
\makeatletter
\newenvironment{nointerfere}[1]
{
	\renewcommand{\assumptionnumber}{Assumption #1$'${\normalfont~(No interference)}}
	\begin{assumption*}
		\protected@edef\@currentlabel{#1$'$}
	}
	{
	\end{assumption*}
}
\makeatother

\newtheorem*{example*}{\examplenumber}
\providecommand{\examplenumber}{}


\def\mysc{SC}
\newcommand{\ind}{\perp\!\!\!\perp}

\def\it{{it}}

\def\donor{\mathcal{D}}

\def\uu{\mu} 
\def\ll{\lambda}
\def\cc{C}  
\def\sumi{\sum_\ii}
\def\ee{\varepsilon}
\def\transpose{{\sf \scriptscriptstyle{T}}}
\def\trans{^{\transpose}}
\def\inv {^{\scriptscriptstyle -1}}

\def\coefc{\xi}
\def\what{\widehat{\alpha}} 
\def\w{\alpha}

\def\gmmweight{\Omega}
\def\ii{{i\in \donor}}
\def\jj{{j\in [N]\backslash\donor}}
\def\ww        {  {W_{ it,\ii}}  }
\def\supww  {  {W_{ jt,\jj}}  }
\def\bw        {  {\w_{\scriptscriptstyle i,\ii}}  }  
\def\bwhat   {  {\what_{\scriptscriptstyle i,\ii}}  }
\def\eedonor{  {\ee_{\scriptscriptstyle i,\ii}}   }
\def\eesup    {  {\ee_{\scriptscriptstyle j,\jj}}  }

\def\myx{\mathcal{V}}
\def\myz{\mathcal{D}}

\def\Sigmaxz{E[\myx_t\myz_t\trans]}

\begin{document}
	
	\def\spacingset#1{\renewcommand{\baselinestretch}
		{#1}\small\normalsize} \spacingset{1}
	
	\date{}{}

	\if1\blind
	{
		\title{\bf Theory for Identification and Inference with Synthetic Controls: A Proximal Causal Inference Framework}
		\author{Xu Shi$^{1*}$, Kendrick Qijun Li$^{1*}$, Wang Miao$^2$, Mengtong Hu$^1$, \\and Eric Tchetgen Tchetgen$^3$ \\
			$^1$Department of Biostatistics, University of Michigan
			\\$^3$ Statistics Department, The Wharton School, University of Pennsylvania\\
			$*$ indicates co-first author}
		\maketitle
	} \fi
	
	\if0\blind
	{
		\bigskip
		\bigskip
		\bigskip
		\begin{center}
			{\LARGE\bf Theory for Identification and Inference With Synthetic Controls: A Proximal Causal Inference Framework}
		\end{center}
		\medskip
	} \fi
	
	\bigskip
	\begin{abstract}
		Synthetic control ({\mysc}) methods are commonly used to estimate the treatment effect on a single treated unit in panel data settings. An {\mysc} is a weighted average of control units built to match the treated unit, with weights typically estimated by regressing \ken{(summaries of)} pre-treatment outcomes \ken{and measured covariates} of the treated unit to those of the control units. However, it has been established that in the absence of a good fit, such regression estimator will generally perform poorly. In this paper, we introduce a proximal causal inference framework to formalize identification and inference for both the {\mysc} and ultimately the treatment effect on the treated, \ken{based on the observation that control units not contributing to the construction of an {\mysc} can be repurposed as proxies of latent confounders}. We view the difference in the post-treatment outcomes between the treated unit and the {\mysc} as a time series, which opens the door to various time series methods for treatment effect estimation. The proposed framework can accommodate nonlinear models, which allows for binary and count outcomes that are understudied in the {\mysc}  literature. We illustrate with simulation studies and an application to evaluation of the 1990 German Reunification.
	\end{abstract}
	
	\vfill
	
	\newpage
	\spacingset{1.9}
 \tableofcontents
 \clearpage
	\section{Introduction}\label{sec:intro}
	
	Synthetic control ({\mysc}) methods are commonly used to estimate the impact of an intervention when one has observed time series data on a single treated unit and multiple untreated units in both pre- and post- treatment periods \citep{abadie2003economic,abadie2010synthetic}. The treated unit is matched to a weighted average of control units, referred to as an ``{\mysc}'', such that the {\mysc}'s post-treatment outcome predicts the treated unit's unobserved potential outcome under no treatment.
		
		In practice, this is often operationalized by regressing (summaries of) the pre-treatment outcome and covariates of the treated unit on those of the control units using ordinary or weighted least squares (OLS/WLS) under the constraints that the coefficients are non-negative and sum to one, and taking the estimated regression coefficients as {\mysc} weights.
		The treatment effect on the treated unit is then estimated as the difference of post-treatment outcomes between the treated unit and its {\mysc}.
		
		The {\mysc} method has become increasingly popular in recent years with a surge of new methods including methods for multiple treated units  \citep{abadie2021penalized,ben2021synthetic}, matrix completion and matrix estimation \citep{athey2021matrix,bai2021matrix,amjad2018robust}, penalization \citep{doudchenko2016balancing,ben2021augmented}, and statistical inference \citep{li2017statistical,chernozhukov2021exact,cattaneo2021prediction}.

		\setstcolor{red}
		The original {\mysc} method of \cite{abadie2010synthetic} was shown to have good theoretical properties assuming a perfect match can be obtained in the observed data, i.e., there exists a set of weights such that the weighted average of the pre-treatment outcomes and covariates among control units is equal to that of the treated unit.
		 \textcolor{black}{In practice, \cite{abadie2010synthetic}  suggested empirically checking if the characteristics of the {\mysc} and the treated unit are sufficiently matched in the pre-treatment period and recommended not to use an {\mysc} if the fit is poor. In the scenario where there is not a good match to the treated unit, }
		
		an alternative framework widely studied in the {\mysc} literature is to assume \textcolor{black}{a perfect match of the underlying unobserved factor loadings rather than the observed data} \citep{ferman2016revisiting,amjad2018robust,powell2018imperfect,ferman2021properties}.  In this setting, \cite{ferman2016revisiting} showed that the OLS estimates of the {\mysc} weights can be inconsistent. Intuitively, the pre-treatment outcome trajectories of control units are proxies of time-varying latent factors but are measured with error. When such proxies of the latent factors are used as independent variables in OLS/WLS, estimated regression coefficients will generally be inconsistent due to the well-known errors-in-variables regression problem \citep{carroll2006measurement}. 
		To address this problem, \cite{ferman2016revisiting} proposed to use lagged control unit outcomes as instrumental variables to construct unbiased estimating equations. \cite{amjad2018robust} proposed to de-noise the data via singular value thresholding. However, both required that idiosyncratic error terms are independent across units and time.

	In this paper, focusing on the setting where there is a good match of the factor loadings but the pre-treatment fit might be poor, we take a completely different view: we argue that the outcomes of control units are essentially proxies of the latent factors. Rather than directly regressing the outcome on all such proxies, one can split the set of proxies into two, thus leveraging one set of proxies to assist the construction of a {\mysc} defined in terms of the other set of proxies. In fact, generally not all control units contribute to the construction of an {\mysc} in practice. The outcomes of control units not included in the {\mysc} can be repurposed to construct identifying moment equations of {\mysc} weights. Besides outcomes of control units, measured covariates contemporaneous with outcomes of the treated and control units can also serve as candidate proxies. As we show in this paper, leveraging such proxies of latent factors can yield identification of {\mysc} weights under certain conditions, which is motivated by  recent work on proximal causal inference  \citep{miao2018identifying,tchetgen2020introduction}. 
	Under certain identifiable conditions, our methods produce consistent estimators of {\mysc} weights and the treatment effect allowing for serial correlation of error terms, as well as both stationary and non-stationary latent factors \citep{hsiao2012panel,ferman2016revisiting}. We also establish nonparametric identification which relaxes the linear factor model assumption.
   Moreover, we propose to view the difference in the post-treatment outcomes between the treated unit and the {\mysc} as a time series where the treatment effect on the treated unit is captured by a deterministic time trend. This observation opens the door to a rich and extensive literature on time series analysis for estimation of the treatment effect.
	
	Our paper is organized as follows. We review classical SC methods and discuss the potential failure of existing {\mysc} methods under imperfect pre-treatment fit of OLS/WLS estimation, even if a perfect match of latent factor loadings can reasonably be assumed, in Section~\ref{sec:inconsistent}. We then propose a proximal causal inference method for consistent estimation and formal statistical inference in Section~\ref{sec:linear}. 
	\textcolor{black}{Specifically, we first focus on estimation of the {\mysc}  weights  in Section~\ref{sec:identify_est}. Then we propose  time series analysis of the time-varying treatment effects on the treated unit, as well as joint estimation and inference for both {\mysc} weights and deterministic trends describing the treatment effect trajectory on the treated unit in Section~\ref{sec:identify_est2}.
	We then show that the fixed effects model and the corresponding  identification conditions are a special case of a more general identification framework in Section~\ref{sec:nonlinear}. In fact, we establish nonparametric identification of the treatment effect on the treated which accommodates nonlinear models and different types of outcomes such as binary or count outcomes.} We illustrate our proposed methods with simulation studies in Section~\ref{sec:simu} and an application to evaluation of the 1990 German reunification in Section~\ref{sec:application}. Throughout, notions of consistency and valid statistical inference about causal effects are studied under an asymptotic regime with diverging time points, but a fixed number of units.  We close with a brief discussion in Section~\ref{sec:discussion}.
Various extensions of interest are further considered in the Supplementary Material, including a conformal prediction approach grounded in our proximal inference framework in Section A, which may be used to retrospectively infer treatment effects for the treated unit in settings where the post-treatment period is of short to moderate length; and a proximal approach for uniquely identifying and inferring causal effects non-parametrically using proxies in Section E, without necessarily assuming a unique {\mysc}.

	\section{Review of classical synthetic control methods
		\label{sec:inconsistent}}
	Suppose $N+1$ units indexed by $i=0,\dots,N$ are observed over $T$ time periods indexed by $t=1,\dots,T$. Consider a binary treatment that impacts unit $i=0$ after a certain time point. Let $T_0$ and $T_1=T - T_0$ denote the number of pre- and post-treatment periods, respectively. Units $i=1,\dots, N$ are untreated control units. We focus on the scenario where $N$ is fixed and $T_0$ and $T_1$ are both large and of roughly the same magnitude. \ken{In Section A of the Supplementary Material, we consider inference for causal effects where there are only a few post-treatment periods.} Let $Y_t$ and $W_{it}$ denote the observed outcome of the treated unit and the $i$-th control unit, respectively,  at time period $t$. Likewise, let $(Y_t(1),Y_t(0))$ and $(W_{it}(1),W_{it}(0))$ denote the pair of potential outcomes of the treated and $i$-th control units one would have observed if, possibly contrary to fact, treatment had or had not been assigned to the treated unit at time period $t$. We are interested in the average treatment effect on the treated unit (ATT) at time $t$  post-treatment, i.e., $$\tau_t=E[Y_t(1)-Y_t(0)]$$ for any $t\geq T_0$. We assume that the observed outcome is a realization of the potential outcome under the assigned treatment value:
	\begin{assumption}[Consistency]\label{asmp:consistency} For any unit $i$, $Y_t = Y_t(0)$ and $W_{it}=W_{it}(0)$ for $t\leq T_0$ and $Y_t = Y_t(1)$ and $W_{it}=W_{it}(1)$ for $t > T_0$.
	\end{assumption}
	A common {\mysc} model in the literature assumes that the outcomes of the treated and control units follow the data generating mechanism \citep{bai2009panel} below:
	\begin{assumption}[Interactive fixed effects model]\label{asmp:model1} For any unit $i$ and time $t$, 
				\begin{equation}
			\begin{alignedat}{2}\label{model1}
				Y_t &= \left\{\begin{array}{ll} \beta_t+\uu_0\trans\ll_t+ \ee_{0t}&\mbox{if $t> T_0$}\\
			    \qquad\uu_0\trans\ll_t+ \ee_{0t}&\mbox{if $t\leq T_0$}	\end{array}\right.\\
				W_\it &= \qquad\quad\,\,\uu_i\trans\ll_t+ \ee_\it,	 
			\end{alignedat}		
		\end{equation} 
		where $\beta_t$ is the time-varying coefficient of  treatment (assumed to be fixed) at each $t$; $\ll_t\in \mathcal{R}^r$ is an $r\times 1$  vector of unmeasured common factors \textcolor{black}{(assumed to be random but with no distributional restrictions imposed for identification, and thus can be either stationary or non-stationary)}; $\uu_i\in \mathcal{R}^r,~i=0,\dots,N$ is an $r\times 1$  vector of  unit-specific factor loadings (assumed to be fixed); and  
		$\ee_\it,~i=0,\dots,N$ is the error term with
		\begin{equation}\label{asmp:errormeanind}
			E[\ee_\it\mid  \ll_t]=E[\ee_\it]=0 \text{ for all $i$ and $t$.}
		\end{equation}
	\end{assumption}

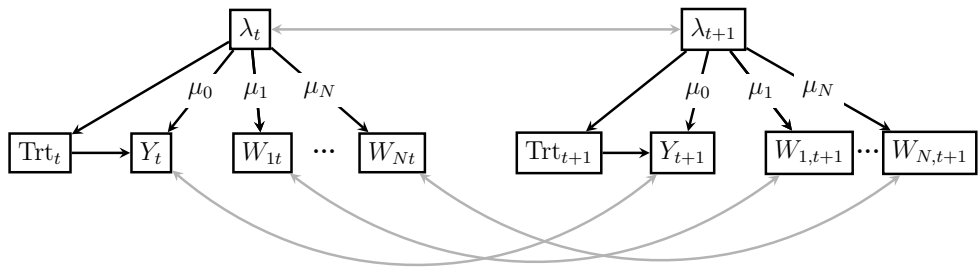
\begin{figure}[!h]
		 		\centering	\resizebox{0.8\textwidth}{!}{
		\begin{tikzpicture}
				\node[draw,line width=0.016in] (U) at (3.5,2) {$\lambda_t$};
				\node[draw,line width=0.016in] (U1) at (11,2) {$\lambda_{t+1}$};
				
				\node[draw,line width=0.016in] (A) at (0.1,0) {$\text{Trt}_t$};
				\node[draw,line width=0.016in] (Y) at (1.9,0) {$Y_t$};
				
				\node[draw,line width=0.016in] (W0) at (3.7,0) {$W_{1t}$};\draw[-stealth,line width=0.016in] (U) -- (W0);
				\node[draw,line width=0.016in] (W) at (5.8,0) {$W_{Nt}$};
				
				\node[draw,line width=0.016in] (A1) at (8.5,0) {$\text{Trt}_{t+1}$};
				\node[draw,line width=0.016in] (Y1) at (10.5,0) {$Y_{t+1}$};
				
				\node[draw,line width=0.016in] (W10) at (12.55,0) {$W_{1,t+1}$};\draw[-stealth,line width=0.016in] (U1) -- (W10);
				\node[draw,line width=0.016in] (W1) at (14.5,0) {$W_{N,t+1}$};
				
				\coordinate [label={\textbf {...}}] (dots1) at (4.7,-0.15);
				\coordinate [label={\textbf {...}}] (dots1) at (13.5,-0.15);

\foreach \from/\to in {A/Y,U/A,A1/Y1,U1/A1}
\draw[-stealth,line width=0.016in] (\from) -- (\to);
{
			\draw[-stealth,line width=0.016in] (U) -- (Y) node[midway,fill=white] {{$\uu_0$}};
				\draw[-stealth,line width=0.016in] (U1) -- (Y1) node[midway,fill=white] {{$\uu_0$}};
				\draw[-stealth,line width=0.016in] (U) -- (W0) node[midway,fill=white] {{$\uu_1$}};
				\draw[-stealth,line width=0.016in] (U) -- (W) node[midway,fill=white] {{$\uu_N$}};
				\draw[-stealth,line width=0.016in] (U1) -- (W10) node[midway,fill=white] {{$\uu_1$}};
				\draw[-stealth,line width=0.016in] (U1) -- (W1) node[midway,fill=white] {{$\uu_N$}};
				}

				\draw[stealth-stealth,color=black!30,line width=0.016in] (U) to[bend left=0] (U1);
				\draw[stealth-stealth,color=black!30,line width=0.016in] (Y) to[bend right=40] (Y1);
				\draw[stealth-stealth,color=black!30,line width=0.016in] (W) to[bend right=37] (W1);
				\draw[stealth-stealth,color=black!30,line width=0.016in] (W0) to[bend right=37] (W10);
		\end{tikzpicture}}
		\caption{\label{fig:DAG_sc} A graphical illustration of the relationships among the treatment status at each time point ($\text{Trt}_t$), the observed outcomes of the treated unit $Y_t$ and the control units $W_\it,i\neq 0$, and the unobserved common factor $\lambda_t$ at each time point $t$. Measured covariates are suppressed for simplicity.}
	\end{figure}

	Note that the treatment status of the treated unit at each time period is not exogenous but rather depends on $\uu_i\trans\ll_t,~i=0,\dots,N$. The interactive effect  $\uu_i\trans\ll_t$ can be viewed as the main source of unmeasured confounding because it is associated with the treatment status and is also predictive of the outcome. Therefore, we also refer to $\ll_t$ as the unmeasured confounder. 
	Figure~\ref{fig:DAG_sc} presents a graphical illustration of our model assumptions.
	\begin{remark}\label{remark1}
		Our proposed method can also  accommodate additive fixed effects (a special case of $\uu_i\trans\ll_t$) with $Y_t(0)= \uu_0\trans\ll_t+\delta_t+ \zeta_0+\ee_{0t}$ and  $W_\it= \uu_i\trans\ll_t+\delta_t+ \zeta_i+\ee_\it,~i=1,\dots,N$,  as considered in \cite{ferman2021synthetic}, where factor loadings of the unknown common factors $\delta_t$ remains constant across units and the unknown intercepts $\zeta_i$ are unit-specific.
	\end{remark}
	Under Assumptions~\ref{asmp:consistency} and \ref{asmp:model1}, we have
	\vspace{-0.2in}\begin{empheq}[left=Y_t{~=\empheqlbrace}]{alignat=2}
		Y_t(0)=&~\uu_0\trans\ll_t+ \ee_{0t}&&~~\text{ when } t\leq T_0,~~ \label{eq:yzero}\\
		Y_t(1)=&~Y_t(0)+\beta_t && ~~\text{ when } T_0<t<T,~~\label{eq:yone}
	\end{empheq}\vspace{-0.45in}
	\begin{equation}
		W_\it=W_\it(1)=W_\it(0)=\uu_i\trans\ll_t+ \ee_\it, \text{ for any }i,t. \label{eq:nco}
	\end{equation}

	By Eq. (\ref{eq:yone}) and the assumption that $\beta_t$ is a deterministic function of time (i.e., non-random), we have, in the classical SC setting, $$\beta_t=Y_t(1)-Y_t(0)=\tau_t.$$ Eq.~(\ref{eq:nco}) implies no interference, i.e., the treatment assigned to the treated unit  does not impact the outcomes of the control units.  In the post-treatment period, $Y_t(1)=Y_t$ is always observed while $Y_t(0)$ is unobserved. Our goal is to find an {\mysc} such that the post-treatment outcome of the {\mysc} can predict the unobserved $Y_t(0)$ of the treated unit.
	
	A key assumption of the proposed {\mysc} methods is that the unmeasured confounding effect on the treated unit can be matched by a weighted average of the unmeasured confounding effect on a set of the control units, referred to as the ``donor pool". 
	Let $\donor$ index the control units selected into the ``donor pool", and let $|\donor|$ denote the number of control units in the donor pool. For a sequence of unit-specific values $a_i\in\mathcal{R}$, $i=1,\dots,N$, let $a_{\scriptscriptstyle i,\ii}$ denote the $|\donor|\times 1$ vector of $a_i$'s for $i\in\donor$, i.e., units in the donor pool; if $a_i\in\mathcal{R}^d$ is a $d$-dimensional vector, then $a_{\scriptscriptstyle i,\ii}$ is a $ d\times |\donor|$  matrix. Similarly, let $\ww$ denote the $|\donor|\times 1$ vector of $W_\it$'s for $i\in\donor$. The  assumption below is key to identification of an {\mysc}, an extension of which to nonlinear setting is provided in Section~\ref{sec:nonlinear}. 

	\begin{assumption}[Existence of synthetic control]\label{asmp:existence}
		There exist a set of weights  $\bw$ with $\w_i\in\mathcal{R}$ such that 
		\begin{equation}
			\uu_0 = \sumi \w_i \uu_i. \label{eq:scexist}
		\end{equation}
	\end{assumption}

As we formalize below in Theorem~\ref{thm:id-att}, the mean potential outcome of the treated unit under no treatment in the post-treatment period is $E[Y_t(0)]=E[\sumi \alpha_iW_{it}]$, proved in Section B of the Supplementary Material.
\begin{theorem}\label{thm:id-att}
    Under Assumptions~\ref{asmp:consistency}-\ref{asmp:existence}, we have $E[Y_t(0)]=E[\sumi \alpha_iW_{it}]$ for any $t$, and the ATT at time $t$ for any $t\geq T_0$ is
    \begin{equation}\label{eq:id-att}\tau_t =E[Y_t - \sumi \alpha_iW_{it}].\end{equation}
\end{theorem}

	\begin{remark} 
\cite{abadie2010synthetic} assume that the weights are non-negative and sum to one, such that the {\mysc} lies in the convex hull of the donors' characteristics. 
		The sum-to-one condition is implied by Assumption~\ref{asmp:existence} when there exist unknown common factors $\delta_t$ with constant factor loadings as discussed in Remark~\ref{remark1}. The non-negative condition is primarily useful for interpretation sake. 
		Because $\bw$ is estimated by regression, when the number of parameters, $|\donor|$, is larger than the sample size, the convex hull assumption can bypass the need for regularization. Relaxations of the convex hull restriction are discussed in \cite{doudchenko2016balancing}. It is important to note that while $\bw$ satisfying Assumption~\ref{asmp:existence} may not be unique, i.e., there may not be a unique {\mysc}, all {\mysc}s satisfying Assumption~\ref{asmp:existence} lead to the same $E[Y_t(0)]$ using our proposed method, for any $t$ as stated in Theorem~\ref{thm:id-att}. As such, the ATT at time $t$ is identified as long as one can identify at least one set of {\mysc} weights $\bw$ satisfying Assumption~\ref{asmp:existence}. \textcolor{black}{Our proposed method does not preclude the possibility to further require {\mysc}s to fall within the convex hull of donor characteristics, which as argued by \cite{abadie2010synthetic} can provide an appealing interpretation.} 
	\end{remark}
	Under Assumptions~\ref{asmp:model1} and \ref{asmp:existence}   we have  that in the pre-treatment period, 
	\begin{equation}\label{eq:ols_bias}
		Y_t
		= \sumi \w_i W_\it + \left(\ee_{0t}-\sumi \w_i \ee_\it\right), \text{ for any }t\leq T_0.
	\end{equation}
	Thus a natural approach to estimate $\bw$ is regression using pre-treatment data, viewing the observed outcomes $W_{it}=\uu_i\trans\ll_t+\ee_t$ as proxies of the unobserved latent factors $\ll_t$ with measurement error $\ee_t$. The following constrained OLS is widely used to construct and {\mysc}
	\begin{equation}\label{eq:what_inconsistent}
		\bwhat=\underset{\alpha_1\geq 0, \sum_{i\in\donor}\alpha_i = 1}{\arg\min}\;\frac{1}{T_0}\sum_{t=1}^{T_0}\left(
		Y_t-\sumi \w_iW_\it 
		\right)^2.
	\end{equation}

	However, the (constrained) OLS weights are generally inconsistent \ken{even when $T_0\rightarrow\infty$}, and thus treatment effect estimates will generally be biased \citep{ferman2021synthetic}.  This is because in Eq.~(\ref{eq:ols_bias}), the residual $\ee_{0t}-\sumi\w_i\ee_\it$ is correlated with the regressor ${\ww}$, therefore the coefficients $\bwhat$ obtained from (constrained) OLS is inconsistent unless $\ee_\it$ is zero, i.e., a noiseless setting.
	The inconsistency of $\bwhat$ \ken{in the large $T_0$ setting} has been pointed out by \cite{ferman2016revisiting} and \cite{powell2018imperfect}. They showed that $\bwhat$ 
	converges to a set of weights that not only attempt to match the latent factors as in Assumption~\ref{asmp:existence}, but also strain to minimize the noise in the synthetic outcome. 

	\textcolor{black}{It is important to note that this result does not necessarily conflict with that of \cite{abadie2010synthetic}, because rather than assuming perfect match of $\uu_i$, they require a good match of (summaries of) the observed outcomes and covariates with the chosen control units such that the right-hand side of Eq.~\eqref{eq:what_inconsistent} is approximately zero. Under certain regularity conditions they established that the bias of the estimated treatment effect is bounded by a function that goes to zero as $T_0$ goes to infinity, or as the maximal variance of $\ee_\it$ goes to zero. This provides theoretical guarantees a user might expect in a setting where a good fit can be obtained in the pre-treatment period. As suggested by  \cite{abadie2010synthetic}, one can empirically verify whether there is a good match to the treated unit that can serve as an {\mysc}, and one should not use an OLS-based {\mysc} if the pre-treatment characteristics of the treated unit and the {\mysc} are not adequately matched. Our approach introduced below complements traditional {\mysc} methods by focusing on the match of underlying population parameters rather than match of the observed characteristics, and as such accommodates an imperfect {\mysc} pre-treatment match. We view this contribution as expanding the scope of applications of {\mysc} methods}.

		\section{A proximal causal inference approach to {\mysc}~\label{sec:linear}}
	In this section, we introduce a proximal causal inference framework for {\mysc}. Our proposal mitigates the inconsistency in the estimation of {\mysc} weights and ATT by leveraging supplemental proxies that are not included in the construction of an {\mysc} for identification of the {\mysc}. We also demonstrate that $E[Y_t(0)]$ may be recovered by a general function of the donors' outcomes and measured covariates which goes beyond linear combination of the control units and extends the linear interactive fixed effects model.
	For ease of exposition, we introduce our framework ignoring measured covariates unless stated explicitely.  Technical details of our approach to incorporate measured covariates is relegated to Section~C of the Supplementary Material.
 
 \subsection{Identification and estimation of synthetic control weights\label{sec:identify_est}}
	Theorem~\ref{thm:id-att} illustrates the pivotal role the {\mysc} weights play in identification of treatment effects.
 As discussed in Section~\ref{sec:inconsistent}, in the presence of latent variables $\ll_t$, direct regression adjustment of ${\ww}$, which are valid proxies of latent variables,  will generally fail to be consistent due to measurement errors $\ee_\it$. 
	However, as we establish below, consistent estimation is possible if one has access to supplemental proxies which, although not used directly to form an {\mysc}, may be used to identify {\mysc} weights.
	Specifically, suppose that additional proxies of $\ll_t$ are available, which are a priori known to be associated with the treated and donor units in the pre-treatment period  only through $\ll_t$. Formally, 
	\begin{assumption}[Existence of proxy]\label{asmp:nce}
		We have observed $Z_t$ such that $Z_t\ind \{Y_t,{\ww}\}\mid \ll_t$ for any $t\leq T_0$.
	\end{assumption}
	We argue that such proxy variables are often available in {\mysc} settings. Assuming independence of the error terms between the donors and the rest of the control units, i.e. $\{\ee_{0t},\eedonor\}\ind\eesup$, then as mentioned in Section~\ref{sec:intro}, a reasonable candidate for proxy variables is the outcome of units excluded from the donor pool, i.e., $Z_t=\supww$, where $[N]=\{1,\dots,N\}$. In practice, Assumption~\ref{asmp:existence} may require only a subset of control units that provide a good match to the treated unit, i.e., not all control units are used to construct the {\mysc}. In fact, it is customary to restrict $\donor$ to units with similar pre-treatment outcome trajectories and covariates. Methods for selection of donors have been discussed in \cite{doudchenko2016balancing}, \cite{chernozhukov2021exact}, and \cite{li2017estimation}. In addition, a control unit is often excluded from $\donor$ if it might  experience  a spillover effect from the treatment on the treated unit or an intervention that conflicts with the control condition under consideration, both of which violate the no-interference assumption in Eq.~(\ref{eq:nco}), but do not invalidate such units as valid proxy $Z_t$.
	For instance,  in the influential tobacco control program studied by \citet{abadie2010synthetic}, 38 out of 50 states were considered but 11 states introducing similar interventions were excluded from the donor pool, and only 5 states ended up forming an {\mysc}. Such data 
  may be repurposed towards identification and thus continue to play an important role as valid proxy variables. Other candidate proxy variables are measured covariates of units in the donor pool that are contemporaneous with $\{Y_t,{\ww}\}$, because they do not have a causal impact on the outcomes but are associated with $\ll_t$ and $X_t$. 
	
	We now introduce our proposed identification and estimation strategy leveraging proxy variable(s) $Z_t$.
	Under Assumptions~\ref{asmp:model1}-\ref{asmp:existence}  we have,  in the pre-treatment period, 
	\begin{equation}\label{eq:bridge}
		E[Y_t \mid \ll_t]= E[\sumi \w_iW_\it  \mid \ll_t],\quad\forall t\leq T_0.
	\end{equation}
	By Assumption~\ref{asmp:nce} we have 
	\begin{equation}\label{eq:bridge2}
		E[Y_t \mid \ll_t,Z_t]= E[\sumi \w_iW_\it \mid \ll_t,Z_t],\quad\forall t\leq T_0.
	\end{equation}
	Marginalizing over $\ll_t$ on each side of (\ref{eq:bridge2}) with respect to $f(\ll_t \mid Z_t)$ gives
	\begin{equation}\label{eq:bridge3}
		E[Y_t \mid Z_t]= E[\sumi \w_iW_\it  \mid Z_t],\quad\forall t\leq T_0.
	\end{equation}
	Eq.~(\ref{eq:bridge3}) indicates that $\bw$ can in fact be identified upon re-framing the approach in terms of proxies. 	
	We now formalize the above results for {\mysc} weights.
	\begin{theorem}[Moment condition for $\bw$]\label{thm:id-alpha}
		Under Assumptions~\ref{asmp:consistency}-\ref{asmp:nce}, the {\mysc} weights $\bw$ satisfies the moment condition $E[Y_t -\sumi \w_i W_\it  \mid Z_t]=0$ for any $t\leq T_0$.  
	\end{theorem}

    Theorems~\ref{thm:id-att} and \ref{thm:id-alpha} establish a connection between {\mysc} methods and the recently proposed proximal causal inference (PI) framework \citep{miao2018identifying,miao2018confounding,tchetgen2020introduction,cui2020semiparametric}.  Acknowledging that measured covariates are often imperfect proxies of the underlying unobserved confounding factor in practice, the PI framework leverages such proxies to to identify causal effects without necessarily invoking a no unmeasured confounding assumption. The proposed formulation of the {\mysc} framework can be viewed as a special case of the more general proximal causal inference framework with  $W_\it$ and $Z_t$ serving as proxies of the latent factors $\ll_t$.

	Based on the above identification results, we   propose to estimate $\bw$ using pre-treatment data with the following estimating function
	\begin{equation}\label{eq:moment}
		U_t(\bw)=g(Z_t)\left(
		Y_t-\sumi\w_iW_\it
		\right),~t=1,\dots,T_0,
	\end{equation}
	where $g(\cdot)$ is a $d$-dimensional ($d\geq |\donor|$) vector of user-specified functions. We show that 
	\begin{equation}\label{eq:id-h}
 E[U_t(\bw)]=0\end{equation}
 for any $t\leq T_0$ in Section~D of the Supplementary Material. The above moment equation motivates using generalized method of moments (GMM) with $U_t(\bw)$ as the identifying moment restrictions~\citep{hansen1982large}. The GMM solves
 $$\widehat\alpha_{i,i\in\donor}=\underset{\bw}{\arg\min}\;m(\bw)\trans\Omega m(\bw)$$
 where $m(\bw)=1/T_0\sum_{t=1}^{T_0}U_t(\bw)$ is the pre-treatment sample moments evaluated at  $\bw$ and $\Omega$ is a $d\times d$ user-specified symmetric and positive-definite weight matrix. 
	\begin{remark}\label{remark4}
		\textcolor{black}{A sufficient condition for a unique solution to Eq.~\eqref{eq:id-h} is that the $d\times |\donor|$ matrix $E[g(Z_t)\ww]$ is full row rank for any $t\leq T_0$. As the existence of multiple solutions may complicate inference, hereafter we assume the full-rank condition which guarantees unique identification of $\bw$. We refer the readers to \citet{shi2020multiply} when multiple sets of {\mysc} weights may exist in the categorical data setting.  More generally, one may define additional criteria for an ``optimal'' set of synthetic control weights. A recent manuscript by a subset of the authors, formally studying such a strategy in a proximal causal inference setting is adapted to {\mysc} setting in Section E of the Supplementary Material.} 
	\end{remark}

	\begin{example} \label{example1}
		As discussed above, when available, the outcome of supplemental control unit(s) not in the donor pool, $\supww$ may be used as proxy variables,  assuming conditionally independent errors $\{\ee_{0t},\eedonor\}\ind\eesup\mid\lambda_{t}$ for all $t$. Then our proposed method entails estimating $\bw$ based on the following estimating function
		\begin{equation*}
			U_{\text{PI},t}(\bw)=g\left(\supww\right)\left(
			Y_t - \sumi \w_i W_\it
			\right),~t=1,\dots,T_0.
		\end{equation*}

		Importantly, the residual, $Y_t - \sumi \w_i W_\it=\ee_{0t}-\sumi \w_i \ee_\it$, is independent of $\supww$ given $\ll_t$  in the pre-treatment period, hence one can show that $U_{\text{PI},t}(\bw)$ is an unbiased estimating function with mean zero. 
		In contrast, the moment function used in the classical OLS-based {\mysc} method is the standard normal equation 
		\[U_{\text{OLS},t}(\bw)={\ww} \left( Y_t - \sumi \w_i W_\it \right),~t=1,\dots,T_0,\] which does not have mean zero and therefore fails to be unbiased because ${\ww}$ is correlated with the residual. 
	\end{example}

	\subsection{Estimation and inference of the causal effect on the treated
		\label{sec:identify_est2}}
	At each time point $t$, only a single realization of $(Y_t,{\ww})$ is available, which is clearly insufficient to consistently estimate $\tau_t$ via Theorem~\ref{thm:id-att} and perform inference. Therefore, \ken{we propose to estimate treatment effects using parametric models under the setting where both numbers of pre- and post-treatment periods are large and diverge at the same rate, i.e. $T_0,T_1\rightarrow\infty$ and $T_0/T_1\rightarrow\rho\in(0,\infty)$ as $T\rightarrow\infty$, where $\rho$ is a fixed constant. In such a setting, it is possible to pool information across time via some form of smoothing to infer a deterministic trend, e.g., a parametric model encoding smooth dependence on $t$.} For example, when appropriate, one may assume that $\tau_t=\tau(t/T; \gamma)$ is a function of time indexed by a finite-dimensional parameter $\gamma\in\mathcal{R}^k$, such as $\tau_t=\gamma_0+\gamma_1t/T$.

	More generally, $\tau_t$ can be estimated  quite flexibly using standard time series analysis. To see this, we define the following post-treatment contrast for $t\leq T_0$ 
	$$e_t = Y_t - \sumi\w_iW_\it = \tau_t+r_t,$$ 
	which may be viewed as a standard time series where  $\tau_t$ captures deterministic trends over time, including potential secular and seasonal patterns, and $r_t=\ee_{0t}-\sumi\w_i\ee_\it$ is a mean zero residual process.  For estimation and inference, we will assume that the residual process $\{r_t\}_{t=T_0+1}^T$ is stationary and weakly dependent across time periods. Formally,
\begin{assumption}\label{asmp:stationary} The process $r_t$ is stationary (i.e., identically distributed across time periods) and weakly dependent process (i.e., asymptotically uncorrelated with $corr(r_t,r_{t+h})\overset{h\rightarrow\infty}{\rightarrow}0$).
    
\end{assumption}

 \begin{remark}\label{remark3}
  A sufficient condition for Assumption~\ref{asmp:stationary} is that $\epsilon_t$ and $\epsilon_{it}$ for each unit $i$ are all stationary and weakly dependent processes. Previous literature such as the lagged-outcome method of \citet{ferman2016revisiting} assumed that the error terms $r_t$ are independent both across units and over time to ensure identification and consistent estimation. \citet{abadie2010synthetic} indicate that though their finite sample bound on the bias predicated on good fit relies on independent errors, their bias bounds result could in principle be extended to cases with time-dependent errors, $\epsilon_{it}$. For example, as mentioned by a reviewer, under an MA(1) model for $\epsilon_{it}$, it would presumably be enough to exclude the outcome of the last period before the intervention from the variables fitted by the {\mysc}. Our analysis using proximal causal inference allows for dependence of errors across donor units and over time. In particular, we allow $\epsilon_{it}$ to be serially correlated over time, e.g. $MA(1)$ and $AR(1)$ time series. Therefore, our weakly dependent assumption replaces the assumption of independent and identically distributed errors while ensuring that ergodic laws of large numbers and central limit theorems apply.
\end{remark} 

	Given a consistent estimator $\bwhat$ based on pre-treatment data using the methods described in Section~\ref{sec:identify_est}, one can estimate $e_t$ with $\widehat{e}_t = Y_t - \sumi\what_iW_\it$ in the post-treatment period, which can in turn be analyzed as time series data. For example, one may first inspect the estimated time series $\widehat{e}_t$ to determine an appropriate functional form for $\tau(t/T; \gamma)$, which may then be used to estimate $\gamma$ by standard regression techniques. Alternatively, one may estimate $\tau_t$ by standard time series smoothing techniques such as moving averages. When both secular and seasonal pattern exist, one may estimate both components, say using an autoregressive–moving-average (ARMA) model or possible generalizations of the latter with the goal of producing stationary residuals, a process sometimes referred to as pre-whitening a time series. Nonparametric regression methods such as series estimation (e.g. smoothing splines, polynomials, wavelets) for estimation of $\tau_t$ have also been proposed in~\citet{dong2018additive}. As there exist a rich and extensive literature on such standard time series modeling techniques, we do not further discuss them here, and instead refer the readers to the monograph \cite{hamilton2020time} for more detailed exposition of such techniques. \ken{ Effect estimation by viewing $e_t$ as a time series has also been considered by \citet{abadie2003economic}, in which a smooth treatment effect curve was fit using moving averages of $e_t$  to inspect the economic costs of terrorist activities for the Basque Country.}
	
	To illustrate, we focus on the special case of $\tau_t=\tau$, i.e., the causal effect is constant over time, which we refer to as the constant average treatment effect on the treated (CATT). If the causal effect is not constant, CATT estimates the (weighted) average treated effect on the treated over the post-treatment period. Based on Theorem~\ref{thm:id-att}, we propose to estimate $\tau$ using all post-treatment data with
	\begin{equation}
		\label{eq:EEforbeta}
		\widehat{\tau} = \frac{1}{\sum_{t=T_0+1}^T v_t}\sum_{t=T_0+1}^T v_t\left(Y_t - \sumi\what_iW_\it\right)
	\end{equation}
	for user-specified weights $v_t$. The most commonly used weight is equal weight, i.e., $v_t=1$, in which case $
	\widehat{\tau} = \sum_{t=T_0+1}^T \{Y_t - \sumi\what_iW_\it\}/T_1$.

	We note that the above two-stage estimation of $\bw$ using pre-treatment data followed by $\tau$ using post-treatment data can be achieved by stacking Eqs. (\ref{eq:moment}) and (\ref{eq:EEforbeta}) into a combined estimating function for simultaneous estimation of both parameters. We illustrate in the scenario when $v_t=1$.
	Let  $\myz_t=\{{\ww}\trans,\mathbbm{1}(t> T_0)\}\trans\in\mathcal{R}^{|\donor|+1}$, , $\myx_t=\{\mathbbm{1}(t\leq T_0)g(Z_t)\trans,\mathbbm{1}(t>T_0)\}\trans\in \mathcal{R}^{d+1}$, and $\theta=(\bw\trans,\tau)\trans$, then  $\widetilde{U}_t(\theta)=\myx_t\trans(Y_t-\myz_t\trans\theta)$. That is, 
	\begin{equation}\label{eq:EEjoint}
		\widetilde{U}_t(\theta)=
		\begin{pmatrix}
			\mathbbm{1}(t\leq T_0)U_t(\bw)\\
			\mathbbm{1}(t> T_0)(Y_t-\myz_t\trans\theta)
		\end{pmatrix}=
		\begin{pmatrix}
			\mathbbm{1}(t\leq T_0)g(Z_t)\\
			\mathbbm{1}(t>T_0)
		\end{pmatrix}\left(Y_t- \mathbbm{1}(t> T_0)\tau-\sumi\w_iW_\it\right).
	\end{equation}
	We show that $E[\widetilde{U}_t(\theta)]=0$ in Section~D of the Supplementary Material.
	Then we have that the two-stage estimation of $\bw$ followed by ${\tau}$ can be equivalently achieved as
	\[
	\widehat{\theta} = \arg\min_\theta \widetilde{m}(\theta)\trans\gmmweight \widetilde{m}(\theta),
	\]
	where $\widetilde{m}(\theta)=1/{T}\sum_{t=1}^T\widetilde{U}_t(\theta)=1/{T}\sum_{t=1}^T\myx_t(Y_t-\myz_t\trans\theta)$ and $\widetilde \Omega$ is a $(d+1)\times(d+1)$ weighting matrix whose upper-left submatrix equals $\Omega$, lower-right entry equals one, and the rest entries equal zero.
 The resulting estimator $\widehat\theta$ satisfies the following theorem proved in  Section~D of the Supplementary Material.
	\begin{theorem}\label{thm:gmm2}
		Under Assumptions~\ref{asmp:consistency}-\ref{asmp:stationary} with $\tau_t=\tau$,  regularity conditions D.1-D.10 listed in Section~D of the Supplementary Material, and \ken{the condition that the matrix $E[g(Z_t)\ww]$ is full row rank for any $t\leq T_0$,} we have  
		$\sqrt{T}(\widehat{\theta}-\theta)\stackrel{d}{\to}N(0,\Sigma)$, as $T\to\infty$
		
		where 
		$\Sigma = \Sigma_0 S \Sigma_0\trans $, 
		in which  $S=\lim_{T\to\infty}       Var[\sqrt{T}\widetilde{m}(\theta)]$
		is the variance-covariance matrix of the limiting distribution of $\sqrt{T}\widetilde{m}(\theta)$, and $\Sigma_0=(\Sigmaxz\trans\widetilde\Omega\Sigmaxz)\inv\Sigmaxz\trans\widetilde\Omega.$
	\end{theorem}
 In Section D of the Supplementary Meterials, we also included a generalization of Theorem~\ref{thm:gmm2} to incorporate any user-specified parametric models for time-varying treatment effects.
	A consistent estimator of $\Sigma$, denoted as $\widehat{\Sigma}$ can be computed with $\widehat{\Sigma}_0 \widehat{S} \widehat{\Sigma}_0\trans$, where $\widehat{S}$ is a heteroskedasticity consistent (HC) estimator of $S$ when $U_t$ is serially uncorrelated \citep{white1980heteroskedasticity}, and the heteroskedasticity and autocorrelation consistent (HAC) estimator of $S$ when $U_t$ is serially correlated \citep{newey1986simple}, and $\widehat{\Sigma}_0$ is the empirical version of $\Sigma_0$ with the estimate of $\theta$ plugged in.
	
	Both estimation and inference can be implemented with off-the-shelf  R package such as \texttt{gmm}. For instance, one can conveniently call the \texttt{gmm} function of the \texttt{gmm} package as follows

	\texttt{gmm::gmm(g=Y$\sim$X+W,x=$\sim$X+Z)}, where \texttt{Z} denotes the data object for  $\mathbbm{1}(t\leq T_0)*Z_t$,  \texttt{W}   for ${\ww}$,  \texttt{X}   for $\mathbbm{1}(t>T_0)$, and  \texttt{Y}   for $Y_t$. One can further specify whether $U_t$ is serially independent or correlated.
	When $U_t$ is serially independent and homoskedastic, our proposed estimator is equivalent to a simple two stage least squares estimator~\citep{miao2018confounding,tchetgen2020introduction}.
	
	\begin{remark}
		When $Z_t$ satisfies $Z_t\ind(Y_t,{\ww})\mid \ll_t$ for any $t$, a stronger assumption than Assumption~\ref{asmp:nce}, we have that under the CATT model (i.e. $\tau_t = \tau$), $E[Y_t \mid Z_t]= E[\mathbbm{1}(t>T_0)\tau+\sumi \w_iW_\it \mid X_t=x_t,Z_t]$
 for any $t$. Therefore, instead of Eq.~(\ref{eq:EEjoint}) which estimates $\bw$ using pre-treatment data and $\tau$ using post-treatment data, one may jointly estimate $\theta$ using 
		$\widetilde{U}_t(\theta)=
		\tilde{g}(Z_t,\mathbbm{1}(t>T_0))
		(
		Y_t-\mathbbm{1}(t>T_0)\tau-\sumi\w_iW_\it
		),~t=1,\dots,T$,
		where $\tilde{g}(\cdot)$ is a $\tilde{d}$-dimensional ($\tilde{d}\geq 1+|\donor|$) vector of user-specified functions. 
		One could enrich $\tilde{g}(\cdot)$ by $\tilde{g}(t,Z_t)$ to allow for flexible parameter estimation using a subset of data within a specified time period.  With appropriate choice of $\tilde{g}$, this framework unifies both two-stage estimation approach ($\tilde{g}=\{\mathbbm{1}(t\leq T_0)g(Z_t)\trans,\mathbbm{1}(t>T_0)\}\trans$) and joint estimation approach  ($\tilde{g}=\tilde g(Z_t,\mathbbm{1}(t>T_0)$).
		The advantage of joint estimation may be particularly notable in case of a short pre-treatment period relative to post-treatment period, as it could lead to efficiency improvements because both pre- and post-treatment data contribute to estimation of $\bw$. Joint estimation can also be achieved in R using 
		\texttt{gmm::gmm(g=Y$\sim$X+W,x=$\sim$X+Z)} where $Z$ is the the data object for $Z_t$ and the rest are the same as before. 
	\end{remark}
	
	\ken{So far, we have focused on a setting where there are many post-treatment time periods so that it is possible to make inference by pooling information over time. When there are only a few post-treatment time periods, \citet{chernozhukov2021exact} proposed a conformal inference approach to perform hypothesis testing and construct prediction intervals for the treatment effects $\eta_t=Y_t(1) - Y_t(0)$ without smoothing. Their method can be readily adapted to the case where the {\mysc} weights $\bw$ are estimated using the proposed proximal inference approach as described in Section A of the Supplementary Material where we construct prediction intervals for the treatment effects.}

	\subsection{Nonparametric identification and estimation\label{sec:nonlinear}}
	The proximal causal inference framework is in fact far more general than the interactive fixed effects model thus far highlighted. It allows for nonparametric identification of causal effects in the sense that one can uniquely express the causal effect as a function of the observed data distribution without necessarily assuming a linear model \citep{miao2018confounding}.
	In this section, we establish conditions for such nonparametric identification for {\mysc}. Our framework allows for nonlinear models that can be applied to e.g., binary or count outcomes, which are currently under-studied in the {\mysc} literature. 
	
	We first extend previous Assumptions~\ref{asmp:model1}-\ref{asmp:existence}.

	In order to establish nonparametric identification we require the no interference assumption previously discussed in Eq~\eqref{eq:nco}:
	\begin{nointerfere}{\ref{asmp:model1}}{}\label{asmp:nco_ignore}  $W_{it}(1) = W_{it}(0)$ for any $i$ and $t$.
 			\end{nointerfere}
	
	We further extend Assumption~\ref{asmp:existence} to the following assumption that the mechanism by which latent factors affect the potential outcome $Y_t(0)$ can be adequately captured  by an unknown function of a sufficiently rich set of proxies ${\ww}(0)$ in the absence of treatment. 
	\begin{confoundingbridge}{\ref{asmp:existence}}{}\label{asmp:npmodel_bridge} There exists a function $h( \ww )$ such that the outcome model for $Y_t(0)$ is equivalent to a model for $h( \ww(0) )$:
		\begin{equation}
			E[Y_t(0)\mid\ll_t]=E[h( \ww(0) )\mid \ll_t],\quad \forall t\geq 1.
			\label{eq:h_nonlinear}
		\end{equation} 
	\end{confoundingbridge}

	Assumption~\ref{asmp:npmodel_bridge} states that in the absence of treatment, the effect of $\ll_t$ on the treated unit can be uncovered by the effect of $\ll_t$ on the donor units encoded by an unknown function $h(\cdot)$, referred to as a confounding bridge function in the proximal causal inference literature \citep{miao2018confounding,tchetgen2020introduction}. 
	We show existence of $h(\cdot)$ under sufficient conditions in Section~F of the Supplementary Material. Below we provide two examples of confounding bridge functions under linear and nonlinear models.
	\begin{example}[Classical setting]  Assumption~\ref{asmp:npmodel_bridge} is a generalization of Assumptions~\ref{asmp:model1}-\ref{asmp:existence}  to the nonparametric setting. 
		Specifically,  Assumptions~\ref{asmp:model1}-\ref{asmp:existence} imply Eq.~(\ref{eq:bridge}), i.e., $E[Y_t(0) \mid \ll_t]= E[\sumi \w_iW_\it(0) \mid \ll_t],$ which is a special case of Eq.~(\ref{eq:h_nonlinear})  with $h( \ww(0)) = \sumi \w_iW_\it(0)$. In this case, identification of the confounding bridge function is equivalent to identification of the {\mysc} weights $\bw$.
	\end{example}

	\begin{example}[Count outcome]
		For count outcomes, it is more natural to assume a multiplicative model, e.g., $E[Y_t(0)\mid \ll_t]=\exp[\uu_0\ll_t]$	for any $t$. Then Assumption~\ref{asmp:npmodel_bridge} holds with $h( \ww (0))=\exp[f({\ww}(0))]$ if $f$ satisfies $E[\exp[f({\ww}(0))]\mid \ll_t]=\exp[\uu_0\ll_t]$ for all $t$. In particular, if $f(\cdot)$ is linear in $W_\it(0)$, then this condition involves the moment generating function and may hold for Normal and Poisson distributed  outcomes.
	\end{example}
	
	Because $Y_t(1)$ is observed  in the post-treatment period,  one only needs to identify $E[Y_t(0)]$ to obtain the ATT at time $t>T_0$. We have the following nonparametric identification result, which is proved in Section G of the Supplementary Material. 
 \begin{theorem}[Nonparametric identification of ATT]\label{thm:id2} 
     Under Assumptions~\ref{asmp:consistency}, \ref{asmp:nco_ignore} and \ref{asmp:npmodel_bridge}, we have $E[Y_t(0)]=E[h(\ww)]$ for any $t$, and therefore the ATT at time $t$ for any $t>T_0$ is
     $\tau_t=E[Y_t-h(\ww)].$
 \end{theorem}
	
	One cannot directly identify $h(\cdot)$ from Eq.~(\ref{eq:h_nonlinear}) because $\ll_t$ is unobserved. Nevertheless, when supplemental proxy variables $Z_t$ satisfying Assumption~\ref{asmp:nce} are available, $h(\cdot)$ can be inferred from observed data, as stated in Theorem~\ref{thm:id-h2} below and proved in Section~G of the Supplementary Material.
	\begin{theorem}[Moment condition for $h(\cdot)$]\label{thm:id-h2}
		Under Assumptions~\ref{asmp:consistency}, \ref{asmp:npmodel_bridge} and \ref{asmp:nce}, for any $t\leq T_0$, the confounding bridge function $h(\cdot)$ satisfies the moment condition \begin{equation}E[Y_t -h(\ww)  \mid Z_t]=0.\label{eq:id-h2}\end{equation} 
	\end{theorem}

    Theorems~\ref{thm:id-att} and \ref{thm:id-alpha} are a special case of Theorems~\ref{thm:id2} and \ref{thm:id-h2} under more stringent model assumptions, i.e.,  Assumptions~\ref{asmp:model1}-\ref{asmp:existence}. \ken{In general, there may exist multiple solutions to Eq.~\eqref{eq:id-h2}. In Section~H of the Supplementary Material, we show under additional assumptions that (i) the distribution of $\ll_t$ is complete with respect to $Z_t$ and that (ii) the conditional distribution $(Y_t(0), W_t(0))\mid\ll_t$ is stationary, any solution to Eq.~\eqref{eq:id-h2} is a confounding bridge function that satisfies Eq.~\eqref{eq:h_nonlinear}. As such, multiple confounding bridge functions might exist and lead to the same $E[Y_t(0)]$. In the presence of multiple solutions of Eq.~\eqref{eq:id-h2}, we present a nonparametric series estimator of the confounding bridge function and the ATT in Section E of the Supplementary Material, based on recent work by a subset of the authors \citet{zhang2022proximal} which is currently under review elsewhere. On the other hand, the confounding bridge function can be uniquely identified if the distribution of $\ww$ is complete with respect to $Z_t$, which we also establish in Section~H of the Supplementary Material.  Unique identification of $h$  significantly simplifies estimation and inference. Suppose a suitable parametric model $h(\ww;\alpha)$ is posited for the confounding bridge function, where $\alpha$ is a finite-dimensional indexing parameter. Then, identification of $\alpha$ can be expressed as the more standard condition: $E[h(\ww;\alpha)-h(\ww;\alpha')\mid Z_t]\neq 0$ with a positive probability for any $\alpha\neq\alpha'$ and $t\leq T_0$.}

 Similar to Section~\ref{sec:identify_est2}, having identified $h( \ww )$, the residual process $e_t = Y_t - h( \ww )= \tau_t + r_t$ $(t=T_0+1,\dots,T)$
may again be viewed as a standard time series where $\tau_t=\tau(t/T)$ is an unknown smooth function of $t/T\in (0,1]$ that captures deterministic time trends, and $r_t=Y_t(0)-h( \ww )$  is a mean zero error term by Theorem~\ref{thm:id2}. Methods in Section~\ref{sec:identify_est2} can still be applied for effect estimation with simple modification.

	\section{Simulation\label{sec:simu}}
	We investigate the finite sample performance of our proposed method under various conditions. We simulate time series data on $N$ control units and one treated unit over $T_0=50,100,\text{ or } 200$ time period pre-treatment and the same time length post-treatment, i.e., $T_0=T_1$. We assume the following data generating mechanism
 \begin{equation}
			\begin{alignedat}{2}\label{eq:dgp-sim}
				Y_t &= \left\{\begin{array}{ll} \tau+\uu_0\trans\ll_t+ \cc_{0t}\trans\coefc+\ee_{0t}&\mbox{if $t> T_0$}\\
			    \qquad\uu_0\trans\ll_t+ \cc_{0t}\trans\coefc + \ee_{0t}&\mbox{if $t\leq T_0$}	\end{array}\right.\\
				W_\it &= \qquad\quad\,\,\uu_i\trans\ll_t+ \cc_{it}\trans\coefc+\ee_\it,		 
			\end{alignedat}		
		\end{equation}
 where $\tau=2$ for all $t$, 
 $\coefc=1$ or $0$  corresponds to scenarios with or without measured covariates, $\cc_\it\stackrel{i.i.d}{\sim} N(0,1)$ denotes a measured covariate, and
	$\ee_\it\stackrel{i.i.d}{\sim}N(0,1)$ denotes the random errors.
	Under Assumption~\ref{asmp:consistency} we have that $\tau$ is the CATT, i.e., $\tau_t=\tau=2$. We simulate a vector of latent factors $\ll_t=(\ll_{t1},\dots,\ll_{tr})\trans$, where $\ll_{tk}\stackrel{i.i.d}{\sim}N(log(t),1),k=1,\dots,r$, and $r=1, 5$ or $10$. That is, we
	generate three settings with one, five, or ten latent factors. For each setting, we assume the number of control units $N=2r$, and the first half of the control units ($i=1,\dots,r$) constitute the donor pool with $|\donor|=r$. We specify factor loadings $\uu_i,~i=0,\dots,N$ as follows. For the treated unit, the factor loading  is $\uu_0=(1,\dots,1)\trans$. For the control units, 
	the first half of the control units ($i=1,\dots,r$) have factor loadings $\uu_i$ and corresponding {\mysc} weights $\bw$ satisfying Assumption~\ref{asmp:existence}  as follows
	\[
	\begin{blockarray}{c}
		\uu_0 \\
		\begin{block}{(c)}
			1 \\
			\vdots\\
			1\\
		\end{block}
	\end{blockarray}=
	\begin{blockarray}{ccc}
		\uu_1& \cdots & \uu_r  \\
		\begin{block}{(ccc)}
			1 & & \\
			& \ddots & \\
			&  & 1\\
		\end{block}
	\end{blockarray}
	~~\begin{blockarray}{c}
		\bw  \\
		\begin{block}{(c)}
			1 \\
			\vdots\\
			1  \\
		\end{block}
	\end{blockarray} ~\text{ for $r=1,5,10$.}
	\]
	Notably, the {\mysc} weights $\bw$ do not sum to one. The second half of the control units ($i=r+1,\dots,N$) have identical factor loadings as the first half. That is, $\uu_{r+k}=\uu_k$, for $k=1,\dots,r$.

	To estimate $\tau$, we implement our proposed method taking the first half of control units as donors ${\ww}$ and second half of control units as supplemental proxies $\supww$. When there exists a measured covariate (i.e., $\xi=1$) which is predictive of the outcome, we implement our method with and without covariate adjustment to investigate whether there is an efficiency gain from adjusting for such a predictor of the outcome. For inference,  we compute an estimate of HC variance-covariance matrix as detailed in Section~\ref{sec:identify_est2}. 
	
	To compare with the classical {\mysc} method, we implement both  the constrained OLS proposed in \cite{abadie2010synthetic} and the unconstrained OLS, taking all  $N$ control units as donors. 
	For the unconstrained OLS method, we fit the following linear regression model: $Y_t=\mathbbm{1}(t > T_0) \tau+ \sum_{i=1}^N \widetilde{\w}_i W_\it + \cc_0\coefc_i +\nu_\it$, $t=1,\dots,T$, where $\nu_\it$ is some random error. When $\coefc=0$, i.e., when there are no measured covariates, this is equivalent to the classical {\mysc} method without the constraint that $\widetilde{\w}_i\geq 0,i\in\donor$ and $\sum_{i=1}^N \widetilde{\w}_i=1$. When $\coefc=1$, i.e., with measured covariates, this model is an unconstrained version of the estimator proposed by \cite{li2017statistical} for covariate adjustment in {\mysc} methods. 
	For the constrained OLS method, we use the \texttt{Synth} package  \citep{abadie2011synth} with  pre-treatment covariates and outcomes as training data, leaving all hyperparameters at their default value. We extract the estimated {\mysc} weights to predict $Y_t(0)$ in the post-treatment period and then estimate $\tau$. The performance of the constrained OLS in terms of bias is generally worse than the bias of the unconstrained OLS, therefore we only present results for the latter. Simulation results are summarized over 2,000 Monte Carlo samples.

\begin{figure}[h]
		\includegraphics[width=\textwidth]{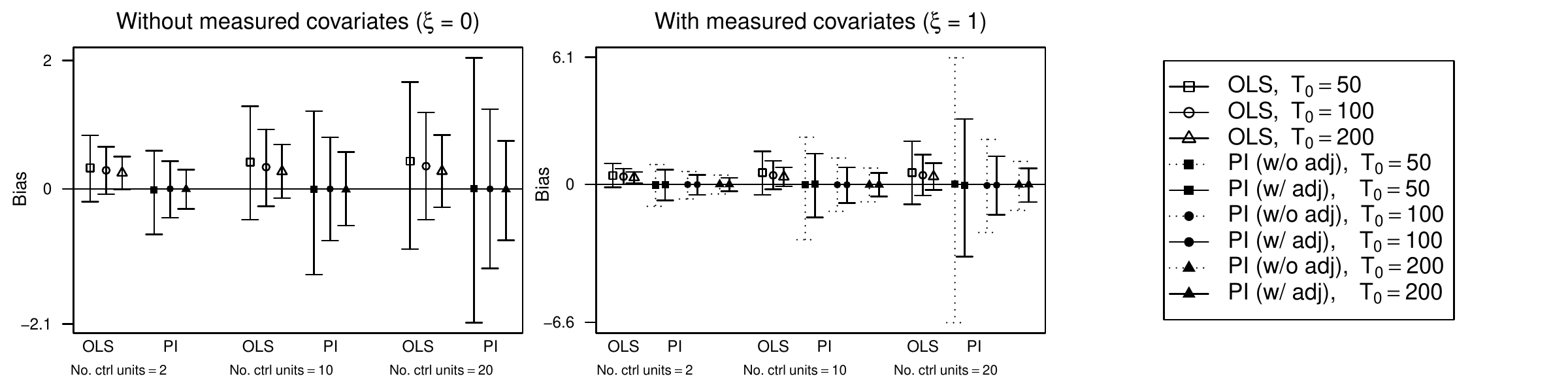}
		\caption{Bias and 95\% Monte Carlo confidence interval of $\tau$ estimates based on the unconstrained regression method (OLS) and our proposed proximal inference method (PI), with a range of number of control units $N=2, 10,\text{ or }20$ and pre- and post-treatment time period $T_0=T_1=50,100,\text{ or } 200$. \label{fig:bias}}
	\end{figure}

	Figure~\ref{fig:bias} presents  bias and 95\% Monte Carlo confidence interval (CI) of the $\tau$ estimates based on the unconstrained regression method (OLS) and our proposed proximal inference method (PI). 
	As expected from theory, the negligible bias of PI approach decreased with increasing $T_0$ and $T_1$. In contrast, the unconstrained OLS regression estimates were substantially biased. This is also consistent with our theoretical expectation, as pointed out in Section~\ref{sec:inconsistent}, that the regression based estimator solves an estimating equation that is biased, and thus does not consistently estimate $\tau$. We also note that such a bias does not decrease as the number of control units increases. This is because, as discussed in \cite{ferman2021properties}, OLS-based weight estimate $\bwhat$ converges to weights that attempt to minimize the variance of $\sumi\w_i\ee_\it$, which does not vanish in our setting as $\bw$ does not decrease with increasing $N$. 
	In addition, our proposed method successfully adjusts for measured covariates with similar bias but higher efficiency as indicated by narrower Monte Carlo CI compared to the PI method without covariate adjustment.
	Table~\ref{tab:cp}   presents Monte Carlo coverage probabilities for PI and unconstrained OLS methods when $\xi=0$ or $1$, with covariate adjustment in the $\xi=1$ scenario. The OLS method had poor coverage across all scenarios, nearly equal to zero for large  $T_0$. In contrast, the PI method had coverage probabilities close to their nominal level of 95\% across all settings considered. 

	\begin{table}[h]
		\caption{\label{tab:cp}
			Coverage probability based on the unconstrained regression method (OLS) and our proposed proximal inference method (PI), with a range of number of control units $N=2, 10,\text{ or }20$ and pre- and post-treatment time period $T_0=T_1=50,100,\text{ or } 200$. 
		}
		\resizebox{1\textwidth}{!}{
			\begin{tabular}{cccccccccccccccc}
				\hline 
				No. & \multicolumn{7}{c}{Without measured covariates ($\xi$ = 0)} &  & \multicolumn{7}{c}{With measured covariates ($\xi$ = 1)}\tabularnewline
				\cline{2-8} \cline{3-8} \cline{4-8} \cline{5-8} \cline{6-8} \cline{7-8} \cline{8-8} \cline{10-16} \cline{11-16} \cline{12-16} \cline{13-16} \cline{14-16} \cline{15-16} \cline{16-16} 
				control & \multicolumn{3}{c}{OLS} &  & \multicolumn{3}{c}{PI} &  & \multicolumn{3}{c}{OLS} &  & \multicolumn{3}{c}{PI}\tabularnewline
				\cline{2-4} \cline{3-4} \cline{4-4} \cline{6-8} \cline{7-8} \cline{8-8} \cline{10-12} \cline{11-12} \cline{12-12} \cline{14-16} \cline{15-16} \cline{16-16} 
				units & 50 & 100 & 200 &  & 50 & 100 & 200 &  & 50 & 100 & 200 &  & 50 & 100 & 200\tabularnewline
				\hline 
				2 & 75.7\%& 66.5\%& 53.7\%&  & 94.8\%& 95.7\%& 95.8\%&  & 65.2\%& 52.2\%& 33.6\%&  & 93.3\%& 94.8\%& 95.1\%\tabularnewline
				10 & 84.6\%& 78.8\%& 75.0\%&  & 96.2\%& 95.7\%& 94.6\%&  & 79.6\%& 74.2\%& 66.7\%&  & 95.6\%& 95.7\%& 96.1\%\tabularnewline
				20 & 88.5\%& 86.4\%& 84.0\%&  & 97.4\%& 95.2\%& 95.7\%&  & 87.4\%& 83.6\%& 78.8\%&  & 98.8\%& 96.7\%& 96.2\%\tabularnewline
				\hline 
			\end{tabular}
		}
	\end{table}
	
	We conduct additional simulation studies under a weakly dependent residual error setting where  $\ee_\it$ is an AR(1) process with coefficient 0.1, i.e., $\ee_\it=0.1\ee_{i,t-1}+\nu_\it$ with $\nu_\it\stackrel{i.i.d}{\sim}N(0,1)$. For PI inference, we report results implementing the HAC estimator of standard errors.  	
	The results are similar to those reported above and are presented in Section~I.1 of the Supplementary Material.
	We further study a time-varying treatment effect setting with $\tau(t/T;\gamma)=\gamma_0+\gamma_1t/T$ where $\gamma=(1,1)\trans$ and we standardize time index $t$ by total time points $T$ to avoid diverging regressors, as recommended by Chapter 16.1 of \cite{hamilton2020time}.
	The results are similar to the above and are presented in Section~I.2 of the Supplementary Material.
	In addition, we investigate a setting where outcomes are Poisson distributed count variables, a setting which, to the best of our knowledge, has not  been formally studied in the  {\mysc} literature. As shown in Section~I.3 of the Supplementary Material, the bias of PI approach again decreased with increasing sample size, with coverage probabilities close to the nominal level. Covariate adjustment tends to be slightly unstable compared to the linear model setting and hence may require relatively larger sample size.

 \ken{The reported bias of constrained OLS in our simulation study is in line with the warning given by \citet{abadie2010synthetic}, \citet{abadie2021using} and \citet{abadie2022synthetic} who recommend use of the constrained OLS estimator only when estimated weights can produce synthetic controls that are a good match for  the treated unit. Note that we did not evaluate the fit in advance in our simulation and found that the results are a good illustration of their warning.}

 \textcolor{black}{Finally, we perform simulation studies to evaluate the conformal inference method for prediction intervals of $\eta_t=Y_t(1)-Y_t(0)$ with the {\mysc} weights estimated using the proposed proximal inference approach. We generate the data under Eq.~\eqref{eq:dgp-sim}, same as before, but set $T_1=1$. We evaluate the average length and coverage rate of the 90\% prediction intervals for $\tau_{T_0+1}$ over 2,000 Monte Carlo samples. We show the results in Section~A.2 of the Supplementary Material. The conformal prediction intervals are well-calibrated in all settings, but appear to be much wider than the ones for the $\tau_t$ estimates above obtained based on data of many post-treatment periods.}

	\section{Application: the 1990 German reunification\label{sec:application}}
	We illustrate the use of our proposed PI method by applying it to a comparative case study of the 1990 German reunification. \cite{abadie2015comparative} studied the effect of the German reunification on per-capita GDP in West Germany using the regression-based {\mysc} method with the additional restriction that the weights are non-negative and normalized to one.
	They collected annual country-level panel data in 1960-2003 for both West Germany, the treated unit, and 16 untreated Organisation for Economic Co-operation and Development (OECD) countries. 
	Data and code to replicate results in \cite{abadie2015comparative} are available online \citep{citedata}.

	The outcome of West Germany, $Y_t$, and the control countries, $W_\it$, is the annual per-capita GDP measured in 2002 U.S. dollars (USD) in country $i$ at time $t$, $i=1,\dots,16,~t=1,\dots,44$. Among the 16 control countries, five (Austria, Japan, Netherlands, Switzerland, USA) were eventually given non-zero weights in \cite{abadie2015comparative} to construct a synthetic West Germany,  while the other 11 states were excluded with zero weights. We hence take outcome trajectories of the five countries as ${\ww}$ and leverage the rest of countries as supplemental proxies, i.e., $Z_t=\supww$. 
	We further adjust for three predictors of GDP, denoted as  $\cc_\it,~i=0,\dots,16$, which include inflation rate (annual percentage change in consumer prices relative to 1995), industry (industry share of value added), and trade openness (measured by export plus imports as percentage of GDP), with missing values imputed via last observation carried forward within each country.
	Other covariates used in \cite{abadie2015comparative} were not employed in our analysis due to excessive missing values. Assuming a linear additive model adjusting for measured covariates, each with unit-specific effects (\ken{Assumption~2'' in Section C of the Supplementary Material}), we estimate $\bw$ and $\coefc$ using the pre-treatment data in 1960-1990 via GMM with a pre-specified identity weighting matrix. We then use the estimated coefficients to compute the synthetic West Germany that aims to replicate the GDP that would have evolved in the absence of the reunification.
	\begin{figure}[htp]
		\centering
		\begin{subfigure}[t]{0.48\textwidth}
			\includegraphics[height=2.5in]{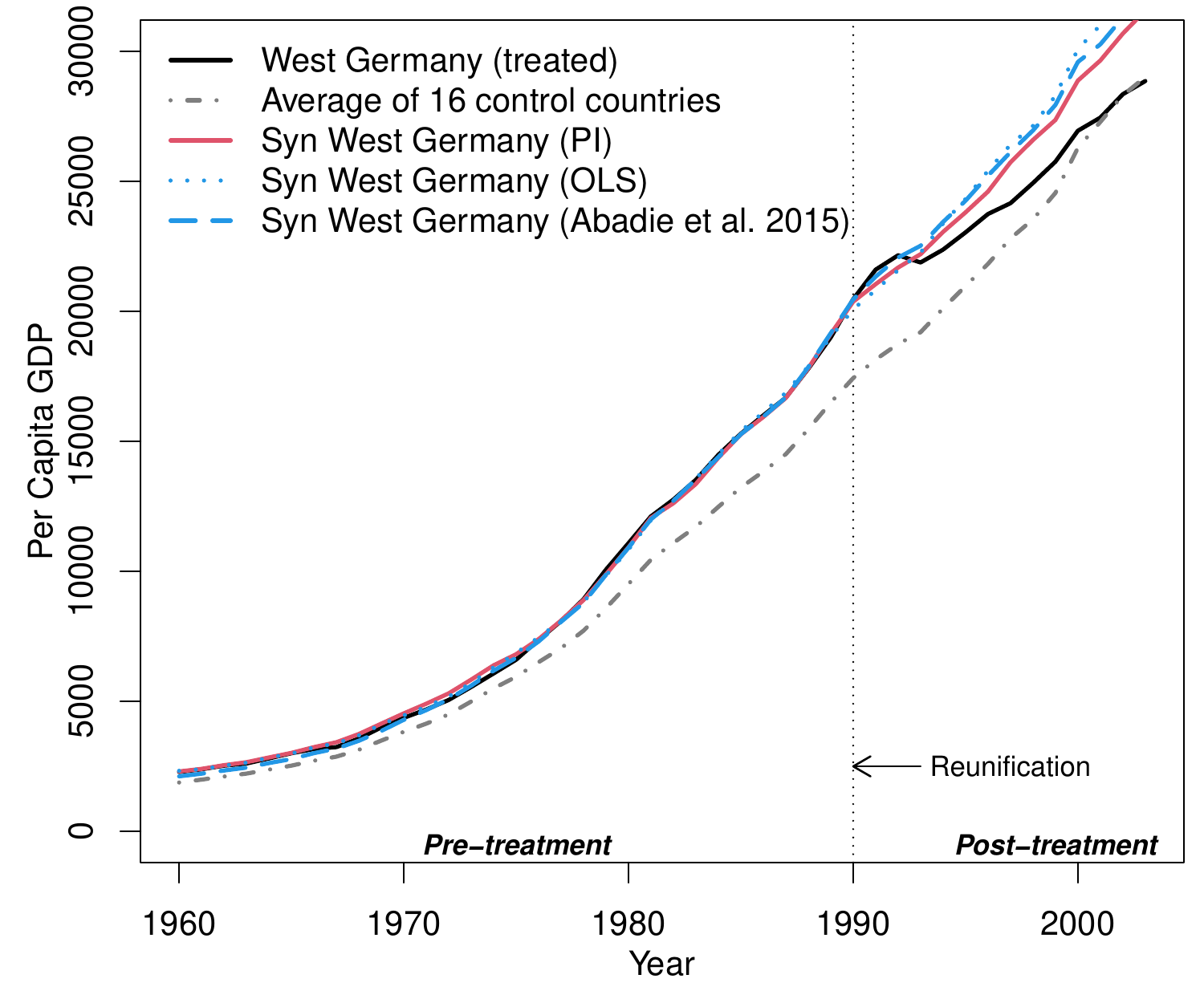}
			\caption{GDP in the pre- and post-treatment period\label{fig:application1}}
		\end{subfigure}
		\begin{subfigure}[t]{0.48\textwidth}
			\includegraphics[height=2.5in]{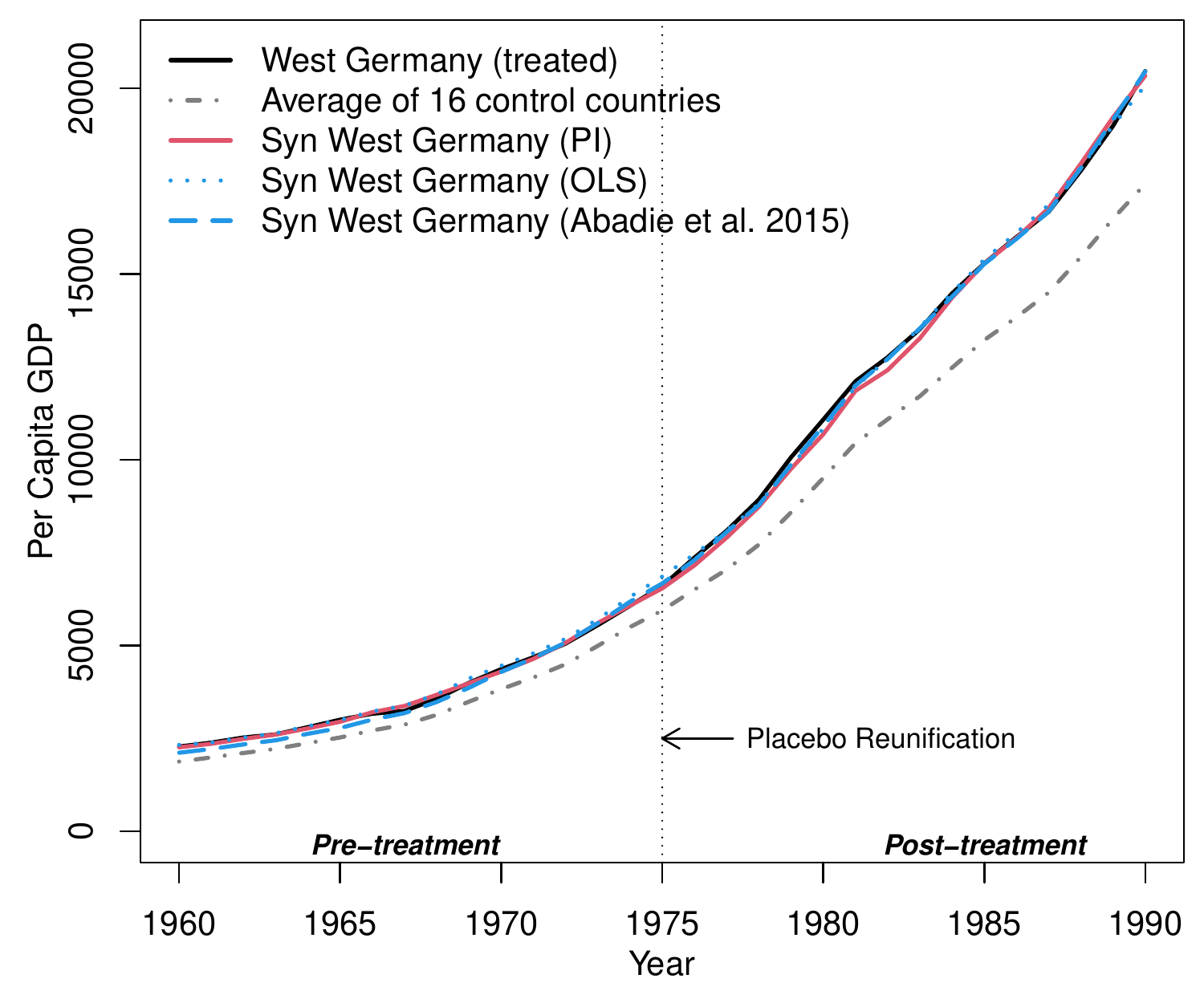}
			\caption{GDP in the placebo period\label{fig:application2}}
		\end{subfigure}
		\caption{Trajectory in per-capita GDP in West Germany and synthetic West Germany. PI: prosimal inference method, OLS: unconstrained OLS regression, Abadie et al. 2015: regression-based method with the additional restriction that the weights are non-negative and sum to one.\label{fig:application}}
	\end{figure} 
	
	Figure~\ref{fig:application} (a) presents the trends of per-capita GDP in West Germany and synthetic West Germany based on (1) a simple average of all 16 control units (2) the PI method, (3) the unconstrained OLS method, and (4) the {\mysc} method in \cite{abadie2015comparative} which further restricts the weights to be non-negative and sum to one.
	In the pre-treatment period, the difference in per-capita GDP comparing  West Germany and  its synthetic counterpart is close to zero from all methods except for the simple average. After 1990 the trends of per-capita GDP diverged.

	\textcolor{black}{We estimate the CATT post reunification using the GMM approach with Eq.~(S.7) in Section C of the Supplementary Material as the estimating function, where $\widetilde g(\widetilde W_{jt,j\in[N]\backslash\donor})=[\mathbbm{1}(t\leq T_0)]\widetilde W_{jt,j\in[N]\backslash\donor}\trans, \mathbbm{1}(t>T_0)]\trans$}. We  inferred that, in the post-reunification period, per capita GDP was reduced, with estimated causal effect  -1200 USD (95\%CI:  -2021,  -380 (HC); -2369,  -32 (HAC)), suggesting a negative impact of reunification on West Germany's GDP. 
	The corresponding estimate in \cite{abadie2015comparative} was somewhat larger, approximately -1600 USD, although formal statistical inference was not available. \textcolor{black}{We also apply the conformal inference approach to construct prediction intervals for $\eta_t = Y_t(1) - Y_t(0)$, causal effects  on the treated unit post reunification. We present the results in Section A.3  of the Supplementary Material.}

	To validate the results, we further conducted a falsification study by restricting estimation to the pre-treatment period, during which the causal effect is expected to be null. We artificially define a placebo-reunification time  of 1975, which falls in the middle of the pre-treatment period. Figure~\ref{fig:application} (b) presents the per-capita GDP trajectories in West Germany and synthetic West Germany computed using the placebo period. 
	\textcolor{black}{The estimated placebo effect was 125 USD (95\%CI:  -289,  539 (HC); -282,  532 (HAC)) using the PI method and -243 USD using the method of \cite{abadie2015comparative}. These results do not provide any evidence invalidating the PI approach.}

	\section{Discussion}\label{sec:discussion}
	In this paper, we have proposed a new framework for  evaluating the impact of an intervention when time series data on a single treated unit and multiple untreated units are observed, in pre- and post- treatment periods. Our proposal is motivated by recent work on proximal causal inference \citep{miao2018confounding,tchetgen2020introduction,cui2020semiparametric}, which is closely related to negative control methods recently developed for analysis of observational data subject to potential unmeasured confounding bias \citep{lipsitch2010negative,shi2020multiply,shi2020selective}. Our proposed framework complements traditional {\mysc} methods in settings with imperfect match of the treated units' and donor units' characteristics, formalizes estimation and inference of the average treatment effect on the treated unit via the generalized method of moments, and also complements the literature on instrumental variables to correct measurement error in panel data~\citep{schennach2016recent}.
	In addition, our observation that the post-treatment outcome differences between the treated unit and the {\mysc} is a time series with treatment effect captured by a deterministic time trend opens the door to various methods on time series analysis for estimation of the treatment effect.
	We further extend the traditional linear interactive effect model to more general cases such as nonlinear models allowing for binary and count outcomes which are rarely studied in the {\mysc} literature. Our methods can  conveniently be extended to accommodate multiple treated units \citep{abadie2021penalized,ben2021synthetic}. A simplest approach might be to do separate PI analyses for each treated unit allowing for different confounding bridge functions and possibly different donor pools. Then unit specific treatment effects could be aggregated in a meta analytic fashion either by assuming, when appropriate, a shared common effect by inverse variance weighted estimation or by allowing for heterogeneous unit-specific effects pooled via random effect approach.

	\ken{We present two approaches for inference of causal effects: the GMM approach and the conformal inference approach inspired by~\citet{chernozhukov2021exact}. The former approach produces inference for $\tau_t$ when there are many post-treatment time points, allowing smoothing over time via a parametric model $\tau(t/T; \gamma)$, while the latter produces inference for  the causal effects at each time point or $\gamma$ when the number of time points in the post-treatment period is small. \citet{cattaneo2021prediction} proposed an alternative approach for obtaining prediction intervals one might potentially consider in combination with our PCI framework. We leave this to future research. }

 \ken{Our work has several limitations. First, in Sections~\ref{sec:linear}-\ref{sec:application}, we do not impose the common restriction on the {\mysc} weights that they are positive and sum to one. We argue that such restrictions are not necessary because we consider the {\mysc} weights that reconstruct the factor loading of the treated unit in the data generating mechanism rather than the observed data. Having said that, incorporating additional restrictions is possible but may result in irregular estimators of the treatment effects~\citep{andrews1999estimation}. Second, in our analysis of German Reunification data, we adjust for measured predictors of GDP. As these measured covariates may also be subject to the impact of reunification, adjusting for them may cause post treatment bias and alter the interpretation of estimated effects. In such cases, methods in mediation analysis may be adapted to estimate alternative causal effect of interest, such as direct effects~\citep{acharya2016explaining,dukes2021proximal}. Third, future research might consider incorporating formal donor selection methods to improve inference and take into account that construction of donor pool may implicitly use information on all units.} 
 
\section*{Supplementary Material}
The Supplementary Material contains proofs and derivations of all technical results presented in the paper and further discussions. 
	\spacingset{1.3} 
	\bibliographystyle{agsm}
	\bibliography{SC}
\end{document}